# The Genetic Code Revisited: Inner-to-outer map, 2*D*-Gray map, and World-map Genetic Representations

H.M. de Oliveira[1] and N.S. Santos-Magalhães[2]

Universidade Federal de Pernambuco
[1] Grupo de Processamento de Sinais
Caixa postal 7.800 - CDU, 51.711-970, Recife-Brazil
[2] Departamento de Bioquímica–Laboratório de Imunologia Keizo-Asami-LIKA
Av. Prof. Moraes Rego, 1235, 50.670-901, Recife, Brazil
{hmo,nssm}@ufpe.br

**Abstract.** *How to represent the genetic code? Despite the fact that it is extensively known, the DNA mapping into proteins remains as one of the relevant discoveries of genetics. However, modern genomic signal processing usually requires converting symbolic-DNA strings into complex-valued signals in order to take full advantage of a broad variety of digital processing techniques. The genetic code is revisited in this paper, addressing alternative representations for it, which can be worthy for genomic signal processing. Three original representations are discussed. The inner-to-outer map builds on the unbalanced role of nucleotides of a 'codon' and it seems to be suitable for handling information-theory-based matter. The two-dimensional-Gray map representation is offered as a mathematically structured map that can help interpreting spectrograms or scalograms. Finally, the world-map representation for the genetic code is investigated, which can particularly be valuable for educational purposes - besides furnishing plenty of room for application of distance-based algorithms.*

## 1. Introduction

**T**he advent of molecular genetic comprises a revolution of far-reaching consequences for humankind, which evolved into a specialised branch of the modern-day biochemistry. In the 'postsequencing' era of genetic, the rapid proliferation of this cross-disciplinary field has provided a plethora of applications in the late-twentieth-century. The agenda to find out the information in the human genome was begun in 1986 [1]. Now that the human genome has been sequenced [2], the genomic analysis is becoming the focus of much interest because of its significance to improve the diagnosis of diseases. Motivated by the impact of genes for concrete goals - primary for the pharmaceutical industry - massive efforts have also been dedicated to the discovery of modern drugs. Genetic signal processing (GSP) is being confronted with a redouble amount of data, which leads to some intricacy to extract meaningful information from it [3]. Ironically, as more genetic information becomes available, more the data-mining task becomes higgledy-piggledy. The recognition or comparison of long DNA sequences is often nearly un-come-at-able.

The primary step in order to take advantage of the wide assortment of signal processing algorithms normally concerns converting symbolic-DNA sequences into genomic real-valued (or complex-valued) genomic signals. *How to represent the genetic code*? Instead of using a look-up table as usual (e.g., [4], [5]), a number of different ways for implementing this assignment have been proposed. An interesting mapping from the information theory viewpoint was recently proposed by Battail [6], which takes into account the unbalanced relevance of nucleotides in a 'codon'. Anastassiou applied a practical way of mapping the genetic code on the Argand-Gauss plane [7]. Cristea has proposed an interesting complex map, termed as tetrahedral representation of nucleotides, in which amino acids are mapped on the 'codons' according to the genetic code [8]. This representation is derived by the projection of the nucleotide tetrahedron on a suitable plane. Three further representations for the 3-base 'codons' of the genetic code are outlined in this paper, namely *i*) inner-to-outer map, *ii*) 2*D*-Gray genetic map, and *iii*) genetic world-chart representations.

## 2. Process of Mapping DNA into Proteins

Living beings may be considered as information processing system able to properly react to a variety of stimuli, and to store/process information for their accurate self-reproduction. The entire set of information of the DNA is termed as the genome (Greek: *ome*=mass). The DNA plays a significant role in the biochemical dynamics of every cell, and constitutes the genetic fingerprint of living organisms [4]. Proteins -consisting of amino acids- catalyse the majority of biological reactions. The DNA controls the manufacture of proteins (Greek: *protos*=foremost) that make up the majority of the dry mass of beings. The DNA sequence thus contains the instructions that rule how an organism lives, including metabolism, growth, and propensity to diseases. Transcription, which consists of mapping DNA into messenger RNA (*m*-RNA), occurs first. The translation maps then the *m*-RNA into a protein, according to the genetic code [3], [4]. Despite nobody is able to predict the protein 3-*D* structure from the 1-*D* amino acid sequence, the structure of nearly all proteins in the living cell is uniquely predetermined by the linear sequence of the amino acids.

Genomic information of eukariote and prokariote DNA is -in a very real sense- digitally expressed in nature; it is represented as strings of which each element can be one out a finite number of entries. The genetic code, experimentally determined since 60's, is well known [5]. There are only four different nucleic bases so the code uses a 4-symbol alphabet: A, T, C, and G. Actually, the DNA information is transcribed into single-stand RNA - the *m*RNA. Here, thymine (T) is replaced by the uracil (U). The information is transmitted by a start-stop protocol. The genetic source is characterised by the alphabet N:={U, C, A, G}. The input alphabet $N^3$ is the set of 'codons' $N^3:=\{n_1,n_2,n_3 \mid n_i \in N, i=1,2,3\}$. The output alphabet $A$ is the set of amino acids including the nonsense 'codons' (stop elements). *A:=*{*Leu*, *Pro*, *Arg*, *Gln*, *His*, *Ser*, *Phe*, *Trp*, *Tyr*, *Asn*, *Lys*, *Ile*, *Met*, *Thr*, *Asp*, *Cys*, *Glu*, *Gly*, *Ala*, *Val*, *Stop*}. The genetic code consequently maps the 64 3-base 'codons' of the DNA characters into one of the 20 possible amino acids (or into a punctuation mark). In this paper we are barely concerned with the standard genetic code, which is widespread and nearly universal.

The genetic code is a map $\mathcal{GC}: N^3 \rightarrow A$ that maps triplets $(n_1,n_2,n_3)$ into one amino acid $A_i$. For instance, $\mathcal{GC}$(UAC)=*Stop* and $\mathcal{GC}$(CUU)=*Leu*. In the majority of standard biochemistry textbooks, the genetic code is represented as a table (e.g. [4], [5]). Let $\|.\|$ denote the cardinality of a set. Evaluating the cardinality of the input and the output alphabet, we have, $\|N^3\|=\|N\|^3=4^3=64$ and $\|A\|=21$, showing that the genetic code is a highly degenerated code. In many cases, changing only one base does not automatically change the amino acid sequence of a protein, and changing one amino acid in a protein does not automatically affect its function.

## 3. The Genetic Code Revisited

Even worthy, ordinary representations for the genetic code can be replaced by the handy descriptions offered in this paper. The first one is the so-called inner-to-outer diagram by Battail, which is suitable when addressing information theory aspects [9]. We present in the sequel a variant of this map by using the notion of the Gray code to systematise the diagram, gathering regions mapped into the same amino acid (Figure 1).

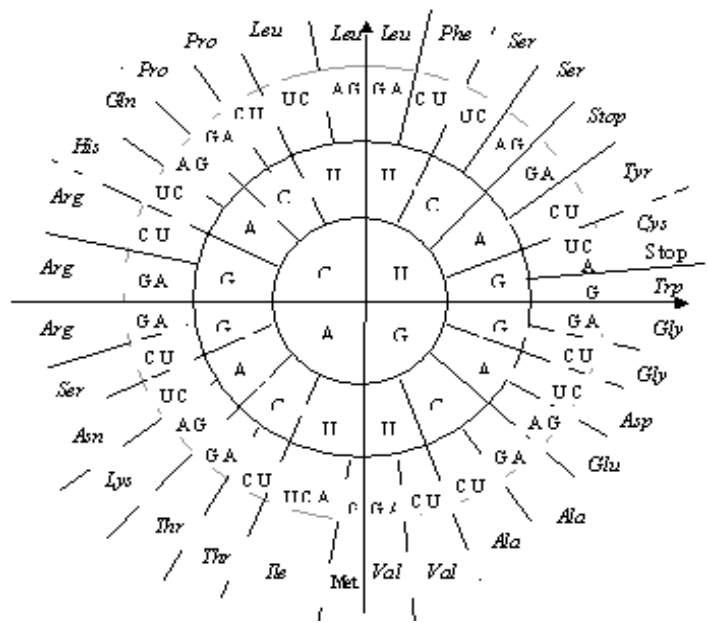

**Fig. 1.** Variant of Battail's inner-to-outer map for the genetic code [6]. The first nucleotide of a triplet (or 'codon') is indicated in the inner circle, the second one in the region surrounding it, the third one in the outer region, where the abbreviated name of the amino acid corresponding to the 'codon', read from the centre, has also been plotted

Another representation for the genetic code can be derived combining the foundations of Battail's map and the technique for generalized 2-*D* constellation proposed for high-speed modems [10]. Specifically, the map intended for 64-QAM modulation shown in Figure 2 can properly be adapted to the genetic code.

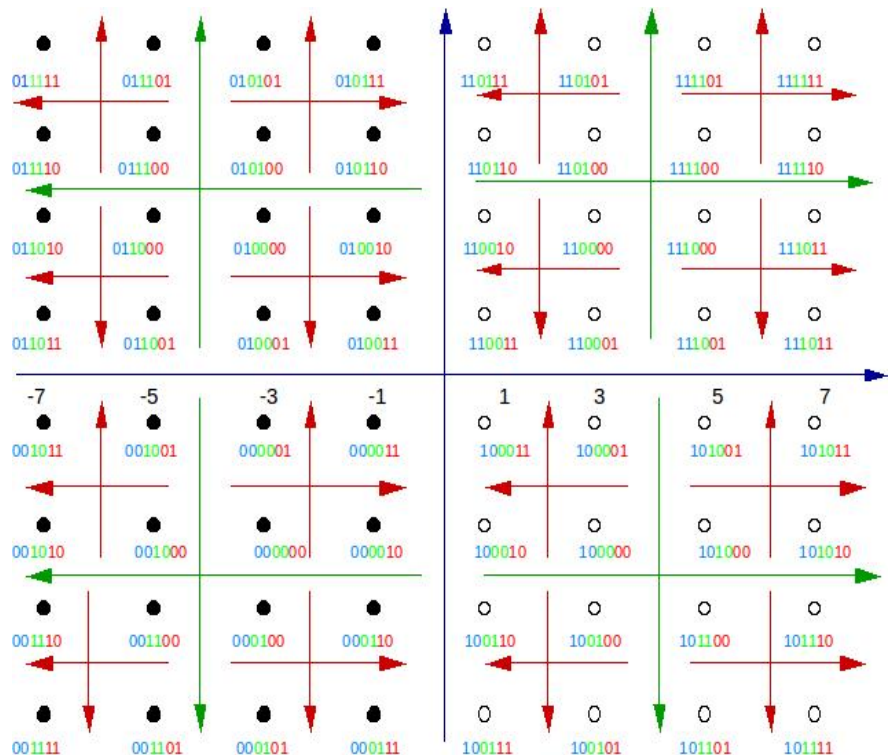

**Fig. 2.** 2*D*-Gray bit-to-symbol assignment for the 64-QAM digital modulation [10].

Binary labels are replaced by nucleotides according to the rule ($x \leftarrow y$ denotes the operator "*replace y by x*"): U ← [11]; A ← [00]; G ← [10]; C ← [01]. The usefulness of this specific labelling can be corroborated by the following argument. The "complementary base pairing" property can be interpreted as a parity check. The DNA-parity can be defined as the sum modulo 2 of all binary coordinates of the nucleotide representations. Labelling a DNA double-strand gives an error-correcting code. Each point of the 64-signal constellation is mapped into a 'codon'. This map (Figure 3a) furnishes a way of clustering possible triplets into consistent regions of amino acids (Figure 3b). In order to merge the areas mapped into the same amino acid, each of amino acids can be coloured using a distinct colour as in Figure 4.

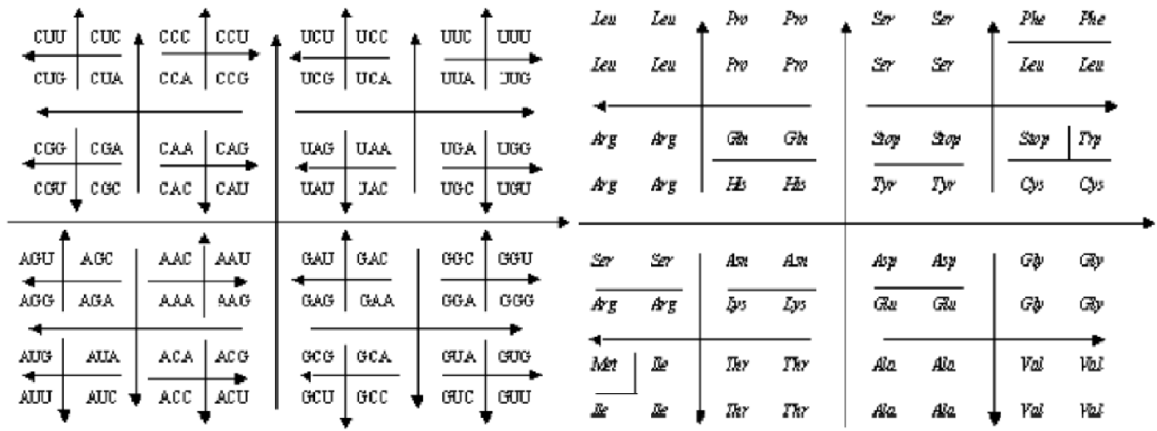

**Fig. 3.** (a) Genetic code map based on the 2*D*-Gray map for the 64 possible triplets ('codons'). Each triplet differs to its four closest neighbours by a single nucleotide. The first nucleotide specifies the quadrant regarding the main axis system. The second one provides information about the quadrant in the secondary axis system, and the last nucleotide (the wobble base) identifies the final position of the 'codon'; (b) Genetic code map based on the 2*D*-Gray genetic map for the 64 possible 'codons' into one of the twenty amino acids (or start/stop)

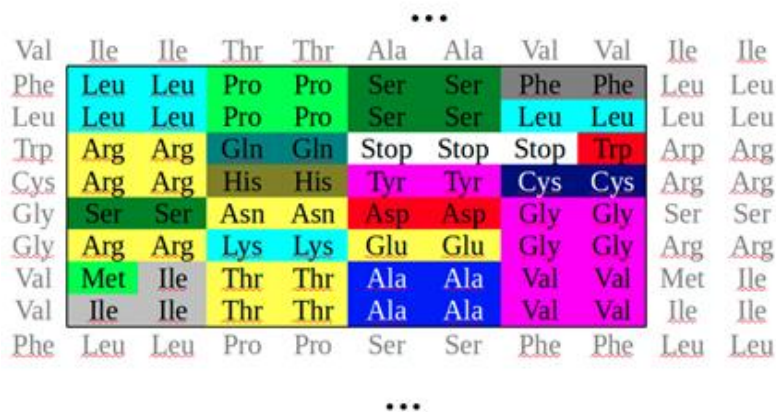

**Fig. 4.** 2*D*-Gray genetic map for the 64 possible 'codons' into one of the twenty possible amino acids (or punctuation). Each amino acid is shaded with a different colour, defining codification regions on the genetic plane. The structure is supposed be 2*D*-periodic

Evoking the two-dimensional cyclic structure of the above genetic mapping, it can be folded joining the left-right borders, and the top-bottom frontiers. As a result, the map can be drawn on the surface of a sphere resembling a *world-map*. Eight parallels of latitude are required (four in each hemisphere) as well as four meridians of longitude associated to four corresponding anti-meridians. The Equator line is imaginary, and the tropic circles have 11.25°, 33.75°, 56.25°, and 78.5° (North and south). Starting from a virtual and arbitrary Greenwich meridian, the meridians can be plotted at 22.5°, 67.5°, 112.5°, and 157.5° (East and west). Each triplet is assigned to a single point on the surface that we named as "Nirenberg-Kohama's Earth"[1] (Figure 5).

---

[1] In honour to Marshall Nirenberg and M. Gobind Kohama, who independently were the main responsible for cracking the genetic code in three nucleotides ('codons')

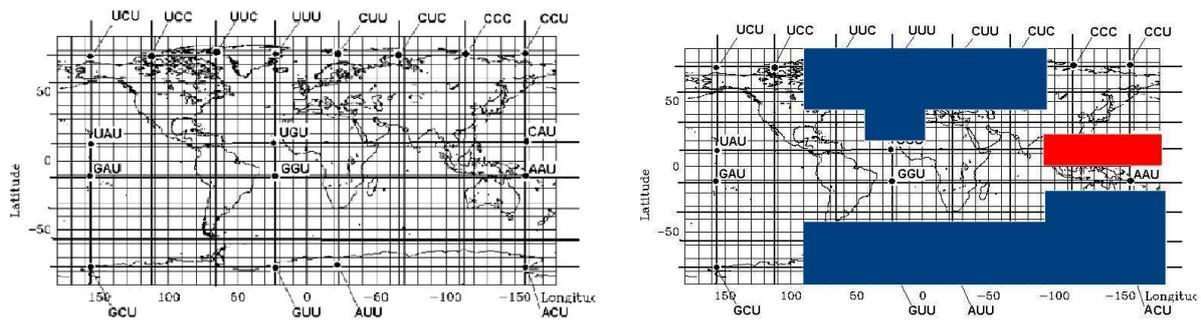

**Fig. 5.** (a) Nirenberg-Kohama's Earth: Genetic code as a world-map representation. There are four meridians of longitude as well as corresponding anti-meridians comprising four 'codons' each. The eight parallels of latitude (tropics) appear as containing eight 'codons' each. The 'codon' UUU, for instance, has geographic co-ordinates (22.5W, 78.75°°N). The Voronoi region [10] of each triplet can be highlighted according to the associated amino acid colour; (b) Continents of Niremberg-Kohama's Earth: regions of essential amino acid correspond to the land and nonessential amino acids constitutes the ocean. There are two continents (one in each hemisphere), and a single island (the Hystidine island)

If every one of essential amino acids are grouped and said to stand for ground, two continents and a lone island emerge (Figure 5b). The remained (nonessential) amino acids materialize the sea. Several kinds of charts can be drawn depending on the criteria used to cluster amino acids [4]. Amino acids can be put together according to their metabolic precursor or coloured by the characteristics of side chains. This approach allows a type of genetic geography. Each one of such representations has idiosyncrasies and can be suitable for analysing specific issues of protein making.

## 4. Closing Remarks[2]

The innovative representations for the genetic code introduced in this paper are mathematically structured so they can be suitable for implementing computational algorithms. Although unprocessed DNA sequences could be helpful, biologists are typically involved in the higher-level, location-based comments on such strings. Much of signal processing techniques for genomic feature extraction and functional cataloguing have been focused on local oligonucleotide patterns in the linear primary sequences of classes of genomes, searching for noteworthy patterns [3], [7]. GSP techniques provide compelling paraphernalia for describing biologically features embedded in the data. DNA spectrograms and scalograms are among GPS powerful tools [7], [11], *which depend on the choice of the genetic mapping*. The miscellany of maps proposed in this paper supplies further cross-link between telecommunications and biochemistry, and can be beneficial for "deciphering" genomic signals. These maps can also be beneficial for educational purposes, furnishing a much rich reading and visualisation than a simple look-up table.

This work was partially supported by the Brazilian National Council for Scientific and Technological Development (CNPq) under research grants N.306180 (HMO) and N.306049 (NSSM). The first author also thanks Prof. G. Battail who decisively influenced his interests